\documentstyle[12pt]{article}

\font\tenmsb=msbm10 scaled\magstep 1
\font\sevenmsb=msbm7 scaled \magstep 1
\font\faivemsb=msbm5 scaled \magstep 1
\newfam\msbfam
\textfont\msbfam=\tenmsb
\scriptfont\msbfam=\sevenmsb
\scriptscriptfont\msbfam=\faivemsb\def\Bbb#1{{\fam\msbfam #1}}

\font\tengothic=eufm10 scaled\magstep 1
\font\sevengothic=eufm7 scaled\magstep 1
\newfam\gothicfam
\textfont\gothicfam=\tengothic
\scriptfont\gothicfam=\sevengothic

\newcommand{\be}{\begin{equation}}
\newcommand{\ee}{\end{equation}}
\newcommand{\dlt}{\delta}
\newcommand{\Dlt}{\Delta}
\newcommand{\ra}{\rightarrow}
\newcommand{\vp}{\varphi}
\newcommand{\bt}{\beta}
\newcommand{\al}{\alpha}
\newcommand{\prt}{\partial}
\newcommand{\Om}{\Omega}
\newcommand{\om}{\omega}

\newcommand{\lbd}{\lambda}
\newcommand{\gm}{\gamma}
\newcommand{\Gm}{\Gamma}
\newcommand{\dgr}{\dagger}

\newcommand{\ep}{\varepsilon}
\newcommand{\bd}{{\bf d}}

\newcommand{\br}{{\bf r}}

\newcommand{\bJ}{{\bf J}}
\newcommand{\bE}{{\bf E}}

\newcommand{\bH}{{\bf H}}
\newcommand{\bA}{{\bf A}}

\begin{document}

\begin{center}
{\Large{\bf Atomic Squeezing under Collective Emission  } \\ [5mm]

V.I. Yukalov$^{1,2,4}$ and E.P. Yukalova$^{3,4}$} \\ [5mm]

{\it
$^1$Bogolubov Laboratory of Theoretical Physics, \\
Joint Institute for Nuclear Research, Dubna 141980, Russia\\ [3mm]

$^2$ Nonlinear Optics Division, Institute of Physics,\\
Adam Mickiewicz University of Pozna\'n, Pozna\'n 61614, Poland \\ [3mm]

$^3$Department of Computational Physics, Laboratory of Information 
Technologies, \\
Joint Institute for Nuclear Research, Dubna 141980, Russia \\ [3mm]

$^4$Institut f\"ur Theoretische Physik, \\ 
Freie Universit\"at Berlin, Arnimallee 14, D-14195 Berlin, Germany}

\end{center}

\vskip 2cm

\begin{abstract}

Atomic squeezing is studied for the case of large systems of radiating
atoms, when collective effects are well developed. All temporal stages are
analyzed, starting with the quantum stage of spontaneous emission, passing
through the coherent stage of superradiant emission, and going to the
relaxation stage ending with stationary solutions. A method of governing
the temporal behaviour of the squeezing factor is suggested. The influence
of a squeezed effective vacuum on the characteristics of collective emission
is also investigated.

\end{abstract}

\vskip 1cm


\newpage

\section{Introduction}

Quantum engineering of squeezed states has been a subject of great
interest during the past several years. The squeezed states of light
[1--3] and atomic squeezing [4--6] have attracted the major attention,
though the effect of squeezing can be defined for any two operators [7].
Atomic squeezing is directly related to the radiation field squeezing
[8] and, vice versa, squeezed atomic states can be created by irradiating
atoms with squeezed light [9--12]. Resonant finite-level atoms are usually
represented by spin operators, because of which atomic squeezing is often
termed spin squeezing. Since the spin operators, employed for describing
finite-level atoms, do not represent real spins, but rather are convenient
mathematical tools, one also uses for atomic squeezing the names of dipole
squeezing [1,4] or pseudospin squeezing [5]. Recently this type of pseudospin
squeezing has been studied for Bose-condensed atoms with two internal states
[13], for two-component mixtures [14], and for multimode Bose-Einstein
condensates [15--17]. There is a variety of possible practical applications
of atomic squeezing, e.g., for atomic spectroscopy and atomic clocks [6], for
atom interferometers [18], and for quantum information processing and quantum
computation [19].

Atomic squeezing for radiating atoms is usually studied by considering a
single atom placed in the so-called squeezed vacuum, that is, subject to the
action of squeezed light. Then one considers the influence of squeezed vacuum
on the radiation process of the atom, whose relaxation characteristics can
be noticeably modified by the presence of the vacuum squeezing [20,21]. An
effective squeezed vacuum can be generated, e.g., by parametric amplifiers
[20,21]. Squeezing of light can be achieved under its interaction with
nonlinear media [22], such as Kerr medium [23]. When light interacts with
condensed matter, there can also arise squeezed exciton-polariton and
phonon-polariton states [24--26] and squeezed magnon states [27]. Squeezed
states of transverse acoustic phonons interacting with light were observed
in experiment [28]. From the theoretical point of view, squeezed polariton
states appear under the canonical Bogolubov transformation diagonalizing the
matter-radiation Hamiltonian [24,29].

If not a single, but several atoms are considered, then one has to take into
account their effective interaction owing to the photon exchange through the
common radiation field. This interaction may also influence atomic squeezing.
The problem of resonance fluorescence from two atoms driven by a squeezed
vacuum field was analysed in Refs. [30--34].

The aim of the present paper is to consider atomic squeezing for an opposite
case, when the number of radiating atoms is large. Then, contrary to the
case of one or two atoms, collective effects come into play.  Owing to photon
exchange, strong interatomic correlations may arise. We aim at studying the
intrinsic atomic squeezing in a multiatomic radiating system. It is especially
interesting to analyze the temporal behaviour of squeezing, starting from
an initially inverted nonequilibrium system, which begins with spontaneous
radiation, passing through the incoherent quantum stage to the stage of well
developed coherence, and then going to the relaxation stage. The evolution
of atomic squeezing through all these stages is the primary concern of the
present paper. Understanding the temporal behaviour of atomic squeezing,
we may find a way for its regulation. One such a way for governing the
evolution of squeezing is suggested.

Another problem, we consider here, is the influence of squeezed vacuum on
the radiation characteristics of a multiatomic system. Such a setup, as is
mentioned above, can be prepared by incorporating the radiating atoms into
a medium with squeezed polariton states. We show that squeezed vacuum can
substantially change the system attenuation.

Thus, we study both sides of the story, the temporal evolution of atomic
squeezing in the presence of collective correlations and the influence on
this evolution of squeezed vacuum.

\section{Atomic squeezing}

In this brief section, we give the main definitions, introducing those
quantities that will be investigated in what follows.

The notion of squeezing can be defined for any two operators, say $A$ and
$B$, which are not necessarily self-conjugate. The definition stems from
the Heisenberg uncertainty relation
$$
\Dlt^2(A) \; \Dlt^2(B) \geq \frac{1}{4}\left |<[A,B]>\right |^2 \; ,
$$
in which the dispersion of an operator is given by
$$
\Dlt^2(A) \equiv \; <A^+ A>\; - \; |<A>|^2 \; ,
$$
with $A^+$ being the Hermitian conjugate of $A$. One says that $A$ is squeezed
with respect to $B$ if
$$
\Dlt^2(A) < \frac{1}{2}|<[A,B]>| \; .
$$
This naturally suggests to introduce the {\it squeezing factor}
\be
\label{1}
Q(A,B) \equiv \frac{2\Dlt^2(A)}{|<[A,B]>|} \; ,
\ee
quantifying how $A$ is squeezed with respect to $B$. If so, then $Q(A,B)<1$.
In terms of factor (1), the Heisenberg relation reads as
$$
Q(A,B)\; Q(B,A) \geq 1 \; .
$$

Resonant atoms are characterized by means of spin operators. For an $i$-th
atom, these are the ladder operators $S_i^\pm$, representing atomic transitions,
and the operator $S_i^z$, representing the population difference. Here and in
what follows, the operators $S_i^\al\equiv S^\al(\br_i,t)$ are treated as the
Heisenberg operators associated with spin $1/2$, which corresponds to two-level
atoms. The Heisenberg equations of motion for these operators translate into
the evolution equations for the related statistical averages. We define the
{\it transition functions}
\be
\label{2}
u_i \equiv 2<S_i^->\; , \qquad u_i^*\equiv 2<S_i^+> \; ,
\ee
{\it local coherence intensity}
\be
\label{3}
w_i \equiv 4|<S_i^\pm>|^2 \; ,
\ee
and the {\it population difference}
\be
\label{4}
s_i \equiv 2<S_i^z>\; .
\ee
The introduced quantities are the functions of space and time, $u_i=u(\br_i,t)$,
$w_i=w(\br_i,t)$, and $s_i=s(\br_i,t)$.

The operators $S_i^\pm$ and $S_i^z$ are the most convenient for characterizing
atomic squeezing. Dealing separately with the operators $S_i^x$, $S_i^y$, and
$S_i^z$, one would meet the so-called trivial squeezing caused by rotation [35].
For the considered operators, one has
$$
\Dlt^2(S_i^\pm) = \frac{1}{2}\; (1\pm s_i) -\; \frac{1}{4}\; w_i \; , \qquad
\Dlt^2(S_i^z) = \frac{1}{4}\; \left ( 1 - s_i^2\right ) \; , \qquad
|<S_i^\pm>| = \frac{1}{2}\; \sqrt{w_i} \; .
$$
The squeezing factor, not involving trivial squeezing [35], and introduced
according to definition (1), writes as
\be
\label{5}
Q(S_i^z,S_i^\pm) = \frac{1-s_i^2}{\sqrt{w_i}} \; .
\ee
One says that the atomic state is squeezed if $Q<1$. From the physical point
of view, this means that, under $Q<1$, the population difference can be measured
with a higher accuracy than the transition characteristics, such as the coherence
intensity (3). To investigate the behaviour of the squeezing factor (5), one has
to know functions (3) and (4).

\section{Collective emission}

Consider a system of $N$ resonant atoms inside matter. The appropriate
Hamiltonian reads as
\be
\label{6}
\hat H = \hat H_a + \hat H_f + \hat H_{af} + \hat H_{mf} \; .
\ee
The Hamiltonian of resonant atoms, with a transition frequency $\om_0$, is
\be
\label{7}
\hat H_a = \sum_{i=1}^N \om_0 \left ( \frac{1}{2} + S_i^z\right ) \; .
\ee
Here and in what follows, we employ the Gaussian system of units and set the
Planck constant $\hbar\equiv 1$. In the field Hamiltonian
\be
\label{8}
\hat H_f = \frac{1}{8\pi} \; \int \left (\bE^2 +\bH^2\right ) \; d\br \; ,
\ee
$\bE$ is electric field and $\bH=\nabla\times\bA$ is magnetic field. For the
vector potential $\bA$, the Coulomb gauge
\be
\label{9}
\nabla\cdot \bA = 0
\ee
will be used. The atom-field interaction is represented by the dipole Hamiltonian
\be
\label{10}
\hat H_{af} = -\sum_{i=1}^N \left ( \frac{1}{c}\; \bJ_i\cdot \bA_i + {\bf P}_i
\cdot\bE_i^0 \right ) \; ,
\ee
in which the transition current is
\be
\label{11}
\bJ_i = i\om_0\left ( \bd S_i^+ -\bd^* S_i^-\right )
\ee
and the transition dipole operator is
\be
\label{12}
{\bf P}_i =\bd S_i^+ + \bd^* S_i^- \; ,
\ee
with $\bd$ being a transition dipole. Here $\bA_i\equiv\bA(\br_i,t)$ is the
vector-potential operator and $\bE_i^0\equiv\bE_0(\br_i,t)$ is an external
electric field. The last term in Eq. (6) represents the matter-field interaction
described by the Hamiltonian
\be
\label{13}
\hat H_{mf} = -\; \frac{1}{c}\; \int {\bf j}_{mat}(\br,t)\cdot\bA(\br,t)\;
d\br \; ,
\ee
where ${\bf j}_{mat}(\br,t)$ is a local density of current in matter.

Writing down the Heisenberg equations of motion, we shall use the standard
commutation relations
$$
\left [ E^\al(\br,t),\; A^\bt(\br',t)\right ] =
4\pi i c\dlt_{\al\bt}(\br-\br') \; ,
$$
in which $c$ is light velocity and
$$
\dlt_{\al\bt}(\br) \equiv \frac{2}{3}\; \dlt_{\al\bt}\; \dlt(\br) - \;
\frac{1}{4\pi}\; D_{\al\bt}(\br)
$$
is the so-called transverse delta-function [22], where
$$
D_{\al\bt} \equiv \frac{\dlt_{\al\bt}-3n^\al n^\bt}{r^3} \; ,
$$
with $n^\al\equiv r^\al/r$ and $r\equiv|\br|$, is the dipolar tensor.

The common way of treating the Heisenberg equations of motion is to take
their statistical averages, with the semiclassical approximation, which results
in the system of the Maxwell-Bloch equations [36]. However, for the present
consideration, it is more useful to follow another approach, by eliminating the
field variables and obtaining the equations containing only the spin variables
[37,38]. This method is, first, more convenient, since for calculating the
squeezing factor (5), we need exactly the averages of the spin variables. And,
second, this approach allows for the consideration of the quantum stage of
collective atomic emission, while the Maxwell-Bloch equations are classical,
thus being unable to treat the quantum stage of evolution.

From the Heisenberg equations of motion for the field variables, using the
Coulomb gauge (9), we have
\be
\label{14}
\left ( \nabla^2 -\; \frac{1}{c^2}\; \frac{\prt^2}{\prt t^2} \right )\;
\bA = -\; \frac{4\pi}{c}\; {\bf j} \; ,
\ee
where the density of current is
\be
\label{15}
j^\al(\br,t) = \sum_\bt \left\{ \sum_{i=1}^N \dlt_{\al\bt}(\br-\br_i)
J_i^\bt(t) + \int \dlt_{\al\bt} (\br-\br') j^\bt_{mat}(\br',t)\; d\br'
\right \} \; .
\ee
The general solution to Eq. (14) writes as
\be
\label{16}
\bA(\br,t) = \bA_{vac}(\br,t) + \frac{1}{c}\; \int {\bf j}
\left ( \br',t -\; \frac{|\br-\br'|}{c}\right ) \;
\frac{d\br'}{|\br-\br'|} \; ,
\ee
where $\bA_{vac}$ is the solution to the related uniform equation, which
represents vacuum fluctuations.

Counting time from the initial moment $t=0$, we keep in mind that all processes
are defined for $t\geq 0$, which can be properly taken into consideration by
the causal condition
$$
S_i^\al(t) = 0 \qquad (t<0) \; .
$$
Also, one should not forget that the very notion of resonant atoms, having
a well-defined transition frequency, has sense only when the interaction of
radiation with an atom is substantially weaker than intra-atomic interactions.
This makes well grounded the Born approximation
\be
\label{17}
S_j^-\left ( t-\; \frac{r}{c}\right ) =
S_j^-(t)\; \Theta(ct-r)\; e^{ik_0r} \; ,
\ee
important for treating the retardation effects in Eq. (16). Here $k_0\equiv
\om_0/c$ and $\Theta(\cdot)$ is a unit-step function.

Taking into account the atomic self-action yields, as is known [39], to the
appearance of the natural width
\be
\label{18}
\gm_0 \equiv \frac{2}{3}\; |\bd|^2 k_0^3 \; .
\ee
This is usually generalized by including in the equations of motion the
longitudinal, $\gm_1\equiv 1/T_1$, and transverse, $\gm_2\equiv 1/T_2$,
relaxation parameters.

The vector potential (16) can be written as a sum
\be
\label{19}
\bA =\bA_{vac} + \bA_{rad} + \bA_{dip} +\bA_{mat} \; .
\ee
Here $\bA_{vac}$ is the vacuum vector potential. The long-range part of the
vector potential generated by radiating atoms is
\be
\label{20}
\bA_{rad}(\br,t) = \sum_j \; \frac{2}{3c|\br-\br_j|} \;
\bJ_j\left (t-\; \frac{|\br-\br_j|}{c}\right ) \; .
\ee
The short-range part is due to the dipolar term resulting in
\be
\label{21}
\bA^\al_{dip}(\br,t) = -\sum_j\; \sum_\bt \;
\int \; \frac{D_{\al\bt}(\br'-\br_j)}{4\pi c|\br-\br'|}\;
J_j^\bt\left ( t - \; \frac{|\br-\br'|}{c}\right ) \; d\br' \; .
\ee
And the last term in Eq. (19) describes the local vector potential produced by
matter,
\be
\label{22}
\bA^\al_{mat}(\br,t) =  \sum_\bt \;
\int \; \frac{\dlt_{\al\bt}(\br'-\br'')}{c|\br-\br'|}\;
j_{mat}^\bt\left (\br'', t - \; \frac{|\br-\br'|}{c}\right )
\; d\br' \; d\br'' \; .
\ee
All short-range parts can be combined into the variable
\be
\label{23}
\xi(\br,t) \equiv 2k_0\; \bd \cdot\left ( \bA_{vac} + \bA_{dip}
+ \bA_{mat} \right ) \; .
\ee

There are in the system two different spatial scales. One is the radiation
wavelength $\lbd\equiv 2\pi c/\om_0$ and another, the mean interatomic
distance $a$. At optical frequencies, one usually has $a\ll\lbd$. The
long-range potential (20) acts at the distance of $\lbd$, while the typical
distance of action for the short-range variable (23) is $a$. Because of the
existence of two different scales, we may employ the scale separation approach
[38,40-42]. Then we separate the operator variables in two parts. One part
is responsible for long-range interactions related to the long-range potential
(20). Since the latter is expressed through the spin operators, we may call
the set
$$
\{ S\} \equiv \{ S_i^+,S_i^-,S_i^z|\; i=1,2,\ldots,N\}
$$
the family of long-range variables. And the set of short-range terms
$$
\{\xi\} \equiv \{ \xi(\br,t)|\; \br\in \Bbb{V}\} \; ,
$$
where $\Bbb{V}$ is the system volume, can be referred to as the family of
short-range variables, or short-range fluctuating fields. Treating $\{ S\}$
and $\{\xi\}$ as two different operator families, we may introduce two types
of averaging for any function $\hat F=\hat F(\{ S\},\{\xi\})$ of these
variables. To this end, we define the statistical averaging over the spin
degrees of freedom as
\be
\label{24}
<\hat F>\; \equiv {\rm Tr}_{ \{ S\} }\;
\hat\rho\; \hat F(\{ S\},\{\xi\}) \; ,
\ee
with $\hat\rho$ being a statistical operator. While the averaging over the
fluctuating fields is denoted as
\be
\label{25}
\ll \hat F\gg \; \equiv {\rm Tr}_{ \{ \xi\} }\;
\hat\rho\; \hat F(\{ S\},\{\xi\}) \; .
\ee

Then we write the Heisenberg equations of motion for the spin operators
$S_i^-=S^-(\br_i,t)$ and $S_i^z=S^z(\br_i,t)$ and average these equations
over the spin degrees of freedom as defined in Eq. (24), which results in
the equations for functions (2), (3), and (4). It is convenient to pass from 
the summation to integration by means of the replacement
$$
\sum_{i=1}^N \Longrightarrow \rho \int d\br \qquad \left ( \rho \equiv
\frac{N}{V} \right ) \; ,
$$
where $\rho$ is the atomic density. In this way, the equations for functions
(2), (3), and (4) are reduced to the equations for the functions $u=u(\br,t)$,
$w=w(\br,t)$, and $s=s(\br,t)$. To write down these equations in a compact
form, we define the effective force
\be
\label{26}
f(\br,t) = f_0(\br,t) + f_{rad}(\br,t) +\xi(\br,t) \; .
\ee
The first term here is due to the external field,
\be
\label{27}
f_0(\br,t) \equiv -2i\bd \cdot \bE_0(\br,t) \; .
\ee
The second term is caused by the radiation potential (20),
\be
\label{28}
f_{rad}(\br,t) \equiv 2k_0\; <\bd \cdot \bA_{rad}(\br,t)>\; .
\ee
And the last term is the fluctuating field (23). The radiation force (28),
taking into consideration Eq. (20), can be presented in the form
\be
\label{29}
f_{rad}(\br,t) = -i\gm_0\rho \; \int \left\{ G(\br-\br',t)\; u(\br',t) -
\; \frac{\bd^2}{|\bd|^2} \; G^*(\br-\br',t)\; u^*(\br',t)
\right \} \; d\br' \; ,
\ee
with the transfer function
$$
G(\br,t) \equiv \frac{\exp(ik_0r)}{k_0r}\; \Theta(ct-r) \; .
$$
As a result, we obtain the equations
$$
\frac{\prt u}{\prt t} = -(i\om_0 + \gm_2)\; u + fs \; ,  \qquad
\frac{\prt w}{\prt t} = -2\gm_2 w + \left ( u^* f + f^* u\right )\; s \; ,
$$
\be
\label{30}
\frac{\prt s}{\prt t} = - \; \frac{1}{2}\; \left ( u^* f + f^* u\right ) -
\gm_1 ( s-\zeta) \; ,
\ee
where $\zeta$ is a stationary population difference for a single atom.
When there is no external nonresonant pumping, then $\zeta=-1$. Generally,
allowing for pumping, one has $\zeta$ in the interval $-1\leq\zeta\leq 1$.
Equations (30) are stochastic differential equations because of the
fluctuating field $\xi$ entering the effective force (26).

The evolution equations (30) can be further simplified in the frame of
the scale separation approach [38,40--42]. For this purpose, we notice
that there are different temporal scales, which follows from the existence
of the small parameters
\be
\label{31}
\frac{\gm_0}{\om_0}\ll 1 \; , \qquad \frac{\gm_1}{\om_0}\ll 1 \; , \qquad
\frac{\gm_2}{\om_0}\ll 1 \; ,
\ee
with the assumption that the atom-field interactions are much weaker than
$\om_0$. Then Eqs. (30) tell us that the function $u$ varies in time much
faster than $w$ and $s$. Hence, the slow functions $w$ and $s$ are temporal
quasi-invariants with respect to the fast function $u$. Therefore, we can
invoke the averaging technique [43] generalized to the case of stochastic
differential equations [38,41,42].

For concreteness, let us consider a sample having typical of lasers
cylindrical shape. The axis of the cylinder is along the $z$-axis, which
is selected by the propagating seed field
\be
\label{32}
\bE_0(\br,t) = \frac{1}{2}\; \bE_1 \; e^{i(kz-\om t)} +
\frac{1}{2}\; \bE_1^*\; e^{-i(kz-\om t)} \; ,
\ee
in which $k\equiv\om/c$, and which is in resonance with the atomic
transition frequency,
\be
\label{33}
\frac{|\Dlt|}{\om_0} \ll 1 \; , \qquad (\Dlt\equiv\om-\om_0) \; .
\ee
In a realistic physical situation, the wavelength $\lbd=2\pi c/\om$ is
much smaller than the radius, $R$, and length, $L$, of the cylindrical
cavity,
\be
\label{34}
\frac{\lbd}{R} \ll 1 \; , \qquad \frac{\lbd}{L} \ll 1 \; .
\ee

When considering a small-aperture cylindric sample, whose Fresnel
number $F\equiv R^2/\lbd L$ is less than one, then there exists the
sole propagating mode and the single-mode description becomes valid.
If the Fresnel number is larger than one, then there arise several
transverse modes, represented by photon filaments [44,45]. The effective
Fresnel number for each filament is less than one. Therefore the problem
again can be reduced to the consideration of separate filaments, for each
of which the single-mode description is applicable [46--49]. In the
single-mode picture, we have
\be
\label{35}
u(\br,t) =  u(t)\; e^{ikz} \; , \qquad w(\br,t) = w(t)\; , \qquad
s(\br,t) = s(t) \; .
\ee

We substitute Eqs. (35) into Eq. (30), multiply the latter by $e^{-ikz}$,
and average over space. We also take into account that the interaction
time $\tau_{int}=a/c$ is very small, since for $a\sim 10^{-8}$ cm, one
has $\tau_{int}\sim 10^{-18}$ s. Therefore after the very short dynamic
interaction stage $0\leq t\leq\tau_{int}$ long-range interactions are
already established. Being so short, this dynamic stage can be neglected,
and one may look for the solutions for $t>\tau_{int}$.

In order to simplify the following expressions, let us introduce the
effective coupling parameters
\be
\label{36}
g \equiv \rho\; \frac{\gm_0}{\gm_2}\;
\int \; \frac{\sin(k_0r-kz)}{k_0r}\; d\br \; ,\qquad
g' \equiv \rho\; \frac{\gm_0}{\gm_2}\;
\int \; \frac{\cos(k_0r-kz)}{k_0r}\; d\br \; ,
\ee
where the integration is over the sample volume $V=\pi R^2L$. These parameters
characterize the effective strength of interaction between the system of atoms
and radiation field. We also introduce the collective width
\be
\label{37}
\Gm \equiv \gm_2( 1 - gs ) \; ,
\ee
collective frequency
\be
\label{38}
\Om \equiv \om_0 + g' \gm_2 s\; ,
\ee
and the dynamic detuning
\be
\label{39}
\dlt \equiv \om - \Om = \Dlt - g' \gm_2 s \; .
\ee

Using the above notation, we can represent the solution for the fast variable
as
\be
\label{40}
u = \left ( u_0 -\; \frac{\nu_1 s}{\dlt+i\Gm}\right ) \; e^{-(i\Om+\Gm)t} +
\frac{\nu_1 s}{\dlt+i\Gm}\; e^{-i\om t} + s \int_0^t \; \xi(t')\;
e^{-(i\Om+\Gm)(t-t')} \; dt' \; ,
\ee
which is obtained from the first of Eqs. (30), keeping fixed all temporal
quasi-invariants, and where
\be
\label{41}
\nu_1 \equiv \bd \cdot \bE_1 \qquad \left ( \frac{|\nu_1|}{\om_0} \ll 1
\right ) \; ,
\ee
and the effective fluctuating field is
\be
\label{42}
\xi(t) \equiv \frac{1}{V} \; \int \xi(\br,t)\; e^{-ikz} \; d\br \; .
\ee

Then we substitute solution (40) into the second and third of Eqs. (30) and
average them over time and over the fluctuating fields, keeping in mind that
the latter are zero-centered, such that $\ll\xi\gg=0$. We define the effective
attenuation
\be
\label{43}
\tilde\Gm \equiv \frac{|\nu_1|^2\Gm}{\dlt^2+\Gm^2}\; \left ( 1  - e^{-\Gm t}
\right ) + \Gm_3 \; ,
\ee
in which the first term is caused by the seed field (32), while the last term
is the quantum attenuation
\be
\label{44}
\Gm_3 \equiv {\rm Re}\; \lim_{\tau\ra\infty}\;
\frac{1}{\tau} \; \int_0^\tau dt \;
\int_0^t \ll \xi^*(t)\xi(t')\gg\; e^{-(i\Om+\Gm)(t-t')} \; dt'
\ee
due to the action of the fluctuating field $\xi(t)$, playing the role of an
effective vacuum. As a result, we obtain the evolution equations for the slow
variables
\be
\label{45}
\frac{dw}{dt} = -2\gm_2(1-gs)\; w + 2\tilde\Gm\; s^2 \; , \qquad
\frac{ds}{dt} = -g\gm_2\; w -\tilde\Gm\; s - \gm_1(s-\zeta) \; .
\ee
These are the final equations defining the functions $w$ and $s$ that we need
to know in order to calculate the temporal behaviour of the squeezing factor
(5).

\section{Stages of evolution}

Equations (45) are rather general and allow one to consider various
situations. Here we concentrate on the case that is, to our mind, the
most interesting, when radiation coherence develops in the system as a
self-organized process. This means that at the initial time no coherence
is imposed upon the atomic system by external forces, but atoms are assumed
to be inverted, which implies the initial conditions
\be
\label{46}
w(0)=0 \; , \qquad s(0)=1\; .
\ee

In order to study the intrinsic atomic squeezing, happening in the radiating
system without a special preparation of a squeezed vacuum, we center, first,
our investigation on the white-noise vacuum, when the fluctuating fields are
correlated as
\be
\label{47}
\ll \xi^*(t)\xi(t') \gg\; = 2\gm_3\dlt (t-t') \; .
\ee
With Eq. (47), we get from Eq. (44) the equality $\Gm_3=\gm_3$.

Aiming at studying a self-organized process, we should not include external
fields acting on atoms. Recall that the role of the seed field (32) has been
to select a resonant atomic mode and to prescribe the cylindric symmetry of
the system. After this has been done, we may set $\nu_1\ra 0$. Consequently,
Eq. (43) yields $\tilde\Gm=\Gm_3=\gm_3$.

As can be checked by straightforward calculations [46--49], for $\lbd\gg a$,
one has $g\gg 1$, where the coupling $g$ is defined in Eqs. (36). Keeping
this in mind, we shall simplify the following formulas by assuming large
$g\gg 1$.

Consider now the relaxation process for the system of inverted atoms, when
the initial conditions (46) are satisfied. Recall that, as is mentioned
above, there always exist a very short interaction stage, when all atoms
radiate independently, not yet feeling each other. This stage lasts in the
time interval
\be
\label{48}
0< t < \tau_{int} \qquad (interaction \; stage) \; ,
\ee
where the interaction time $\tau_{int}=a/c$. Being so short, the interaction
stage can be neglected. The following dynamics is described by Eqs. (45).

In the next stage, atoms begin feeling each other through the photon exchange
of the common radiation field. However atomic phases are not yet correlated,
because of which radiation is incoherent. This is the quantum stage governed
by local quantum fluctuations. The stage lasts in the interval
\be
\label{49}
\tau_{int} < t < t_c \qquad (quantum\; stage)
\ee
till the crossover time
\be
\label{50}
t_c \equiv \frac{T_2}{2g} \; .
\ee
At this stage, the solutions of Eqs. (45), with $\tilde\Gm=\gm_3$, are
\be
\label{51}
w\simeq 2\gm_3 t \; , \qquad
s \simeq 1 - (\gm_1 + \gm_3 - \gm_1\zeta) t \; .
\ee

Coherence in the system develops in a gradual way, being negligible before
the crossover time (50), but becoming important after this time. After the
time $t_c$, the coherent stage comes into play. Equations (45), in the
interval $t_c<t\ll T_1$, yield
\be
\label{52}
w={\rm sech}^2\left ( \frac{t-t_0}{\tau_p} \right ) \; , \qquad
s =\frac{1}{g} - {\rm tanh}\; \left ( \frac{t-t_0}{\tau_p} \right ) \; ,
\ee
where the delay time is
\be
\label{53}
t_0 \equiv \frac{\tau_p}{2} \left ( 1 +\ln\left | 4g\; \frac{\gm_2}{\gm_3}
\right | \right )
\ee
and the pulse time is
\be
\label{54}
\tau_p \equiv \frac{T_2}{g} \; .
\ee
The coherence intensity $w(t)$ reaches its maximum at the delay time (53),
when
$$
w(t_0)=1 \; , \qquad s(t_0) = \frac{1}{g} \; .
$$
Coherent radiation happens in the time interval between $t_c$ and $t_0+
\tau_p$, after which the coherence intensity $w(t)$ exponentially decays.
Hence the coherent stage is characterized by the interval
\be
\label{55}
t_c < t < t_0 +\tau_p \qquad (coherent\; stage) \; .
\ee

The decay of coherence after this stage is described by the solutions
\be
\label{56}
w \simeq 4\exp\left ( - \; \frac{2t}{\tau_p} \right ) \; , \qquad
s\simeq -1 + \frac{2}{g} \; ,
\ee
where $t_0\ll t\ll T_1$. This signifies the beginning of the relaxation
stage, which can be defined as occurring at times
\be
\label{57}
t_0+\tau_p < t < T_1 \qquad (relaxation \;  stage) \; .
\ee
When there is no external stationary pumping, so that $\zeta=-1$, the
relaxation stage is rather trivial. But if there is a strong pumping,
such that $\zeta\approx 1$, the relaxation stage consists of a series of
coherence bursts with gradually diminishing amplitudes. The analysis of
Eqs. (45) shows that the coherent flashing is not periodic though it is
possible to find the approximate time between neighbouring coherent bursts
which is
\be
\label{58}
T_b = \pi\; \sqrt{\frac{2T_1 T_2}{g\zeta} }\; .
\ee
The number of bursts can be estimated as
\be
\label{59}
N_b \sim \frac{T_1}{T_b}  =
\left  ( \frac{g\zeta T_1}{2\pi^2 T_2} \right )^{1/2} \; .
\ee

Finally, for very large time, the process ends with the quasistationary
stage, when
\be
\label{60}
T_1\ll t \qquad (quasistationary\; stage) \; .
\ee
Defining the stationary solutions of Eqs. (45), we keep in mind that, by
definition, $0\leq w\leq 1$ and $-1\leq s\leq 1$. Of possible fixed points, 
we select those that satisfy these restrictions. Generally, when there are 
several fixed points, it is necessary to accomplish the Lyapunov stability 
analysis and to choose the stable fixed point. Following this way, we find 
the stable stationary solutions to Eqs. (45) as
$$
w^* =\frac{1}{2\gm_1\gm_2 g^2} \; \left\{ \gm_1^2(g\zeta -1) +
\gm_3(b-\gm_3) + \gm_1\left [ b -\gm_3(2+g\zeta)\right ] \right \} \; ,
$$
\be
\label{61}
s^* =\frac{\gm_1(1+g\zeta)+\gm_3-b}{2\gm_1 g} \; ,
\ee
in which
$$
b^2 \equiv [ \gm_1(1+g\zeta) + \gm_3]^2 - 4\gm_1^2 g\zeta \; .
$$
These solutions essentially depend on the value of the pumping parameter
$\zeta$. There are three qualitatively different situations: no pumping or
weak pumping $(\zeta\approx -1)$, medium pumping $(\zeta\approx 0)$, and
strong pumping $(\zeta\approx 1)$.

For the case of the medium pumping $(\zeta\approx 0)$, from Eqs. (61), to
third order in $\zeta$, we get
$$
w^* \simeq \frac{\gm_1^2\gm_3\zeta^2}{(\gm_1+\gm_3)^2\gm_2} +
\frac{\gm_1^3\gm_3(\gm_1-\gm_3)g\zeta^3}{(\gm_1+\gm_3)^4\gm_2} \; ,
$$
\be
\label{62}
s^* \simeq \frac{\gm_1\zeta}{\gm_1+\gm_3}\; -\;
\frac{\gm_1^2\gm_3 g\zeta^2}{(\gm_1+\gm_3)^3}\; - \;
\frac{\gm_1^3\gm_3(\gm_1-\gm_3)g^2\zeta^3}{(\gm_1+\gm_3)^5} \; .
\ee

When $|\zeta|\approx 1$, we may simplify Eqs. (61) taking into consideration
that $g\gg 1$. Then to second order in $1/g$, we find
$$
w^*\simeq \frac{(\gm_1+\gm_3)|\zeta|+(\gm_1-\gm_3)\zeta}{2\gm_2 g} \; -
(\gm_1+\gm_3)\;
\frac{(\gm_1+\gm_3)|\zeta|+(\gm_1-\gm_3)\zeta}{2\gm_1\gm_2 g^2|\zeta|} \; ,
$$
\be
\label{63}
s^* \simeq \frac{\zeta-|\zeta|}{2} +
\frac{(\gm_1+\gm_3)|\zeta|+(\gm_1-\gm_3)\zeta}{2\gm_1 g|\zeta|}\; - \;
\frac{\gm_3}{\gm_1 g^2|\zeta|} \; .
\ee
For weak pumping, when $\zeta<0$, this yields
\be
\label{64}
w^*\simeq \frac{\gm_3|\zeta|}{\gm_2 g} \; - \;
\frac{(\gm_1+\gm_3)\gm_3}{\gm_1\gm_2 g^2} \; , \qquad
s^* \simeq \zeta + \frac{\gm_3}{\gm_1 g} + \frac{\gm_3}{\gm_1 g^2\zeta} \; .
\ee
And for strong pumping, when $\zeta>0$, we have
\be
\label{65}
w^* \simeq \frac{\gm_1\zeta}{\gm_2 g} \; - \;
\frac{\gm_1+\gm_3}{\gm_2 g^2} \; , \qquad
s^* \simeq \frac{1}{g}\; - \; \frac{\gm_3}{\gm_1 g^2\zeta} \; .
\ee
These solutions allow us to calculate the behaviour of the squeezing factor
at different stages of the radiation process.

\section{Dynamics of squeezing}

Atomic squeezing is characterized by the squeezing factor (5), which can be
written as
\be
\label{66}
Q = \frac{1-s^2}{\sqrt{w}} \; .
\ee
If the atomic system is prepared in the inverted state, with initial
conditions (46), at the interaction stage (48), one has
\be
\label{67}
Q \simeq 0 \qquad (t\ra 0) \; .
\ee
This means a very strong atomic squeezing.

At the quantum stage (49), using solutions (51), we have
\be
\label{68}
Q \simeq \frac{\gm_1(1-\zeta)+\gm_3}{\sqrt{\gm_3}}\; \sqrt{2t} \; .
\ee
There is yet atomic squeezing, which diminishes with time, since $Q$ grows
with time.

Coherent effects, arising at the coherent stage (55), destroy squeezing. Thus,
at $t=t_0$, the squeezing factor can be approximated as
\be
\label{69}
Q \approx 1 - \; \frac{1}{g^2} \qquad (t=t_0) \; ,
\ee
from where $Q\approx 1$, since $g\gg 1$.

At the relaxation stage (57), the squeezing factor, first, grows as
\be
\label{70}
Q \simeq \frac{2}{g}\; e^{t/\tau_p} \; ,
\ee
and then either monotonely tends to a stationary value, if there is no pumping,
or experiences a series of oscillations, when the pumping is present.

At the quasistationary stage (60), the squeezing factor approaches its
stationary value depending on a particular physical situation. Thus, for the
case of medium pumping $(\zeta\approx 0)$, employing solutions (62), we find
\be
\label{71}
Q \simeq \left [ \frac{\gm_1+\gm_3}{\gm_1|\zeta|} \; - \;
\frac{(\gm_1-\gm_3)g}{2(\gm_1+\gm_3)} \right ] \;
\sqrt{\frac{\gm_2}{\gm_3}} \; .
\ee
Since $|\zeta|\ra 0$, the squeezing factor is large, hence there is no squeezing.

When there is no pumping, or it is very weak, such that $\zeta\approx-1$, then
using Eqs. (64), we get
\be
\label{72}
Q \simeq \left ( \frac{\gm_2\gm_3|\zeta|}{\gm_1^2 g}\right )^{1/2}\;
\left ( 2-\; \frac{1}{g} \right ) \; .
\ee
While if the pumping is strong, that is $\zeta\approx 1$, then  from Eqs. (65),
we find
\be
\label{73}
Q \simeq \sqrt{\frac{\gm_2 g}{\gm_1\zeta}} \; \left [ 1 +
\frac{\gm_1+\gm_3}{2\gm_1 g}\; - \;
\frac{4\gm_1^2+(\gm_1-\gm_3)^2}{8\gm_1^2 g^2} \right ] \; .
\ee
Strong pumping destroys squeezing. If the pumping is absent, so that $\zeta=-1$,
one has atomic squeezing for sufficiently large coupling parameters, when
\be
\label{74}
Q<1 \qquad \left ( g> \frac{\gm_2\gm_3}{\gm_1^2} \right ) \; .
\ee

The temporal behaviour of the squeezing factor (66) can be determined more
accurately by solving numerically Eqs. (45). For this purpose, we consider these
equations for $\tilde\Gm=\gm_3$. To simplify notation, we measure time in units
of $\gm_2^{-1}$, and the relaxation parameters $\gm_1$ and $\gm_3$ in units of
$\gm_2$. Then Eqs. (45) reduce to the form
\be
\label{75}
\frac{dw}{dt} = - 2(1-gs) w + 2\gm_3 s^2 \; , \qquad
\frac{ds}{dt}  = - gw - \gm_3 s - \gm_1(s-\zeta) \; .
\ee
Solving Eqs. (75) numerically, we calculate the squeezing factor (66). The
results are shown in the following figures, in all of which the initial conditions
(46) are assumed.

When the stationary external pumping is absent, so that $\zeta=-1$, the temporal
behaviour of the coherence intensity $w(t)$ and population difference $s(t)$
is qualitatively the same for different parameters $\gm_1$, $\gm_3$, and $g$,
although varying these parameters, of course, modifies the quantitative values
of $w$ and $s$. The typical dependence of $w(t)$ and $s(t)$ on time is presented
in Fig. 1. The related squeezing factor $Q(t)$ is shown in Fig. 2. The coherence
intensity, proportional to the coherent atomic radiation intensity, exhibits the
characteristic superradiant bursts at the delay time $t_0$, when the population
difference falls fast from $s=1$ to almost $s=-1$. The squeezing factor is a
nonmonotonic function of time, whose shape essentially depends on the system
parameters. It may possess a maximum and a minimum, as in Fig. 2, just one
maximum, two maxima and a minimum, as in Fig. 3, or several maxima and minima.

In the presence of an external pumping, when $\zeta=1$, the regime of pulsing 
superradiance evolves. Then the functions $w(t)$ and $s(t)$ display 
oscillations, as is shown in Fig. 4, which results in the appearance of a 
series of oscillations of the squeezing factor in the related Fig. 5. The 
number of these oscillations and their sharpness can be varied by changing 
the system parameters. For instance, the number of oscillations, according to 
Eq. (59), increases with increasing the coupling $g$. This is demonstrated in 
Figs. 6 and 7, which should be compared with Figs. 4 and 5.

Even more possibilities in regulating the temporal behaviour of the squeezing
factor arise in the regime of punctuated superradiance, when the self-organized
evolution of the radiating system is punctuated by short $\pi$-pulses inverting
atoms to their initial inverted state $s=1$. The regime of punctuated spin
superradiance was suggested in Ref. 50. The situation for atomic systems is
analogous to that for spin assemblies. Realizing the regime of punctuated
superradiance for the considered atomic system, we can radically alter the
behaviour of the squeezing factor as is demonstrated in Figs. 8 and 9.

In this way, atomic squeezing in the system of radiating atoms can be regulated
by means of different techniques. One possibility is by preparing a system with
the parameters assuring the desired behaviour of the squeezing factor. Another
method is by organizing the regime of punctuated superradiance, with a chosen
temporal location of inverting pulses. The combination of these methods is also
admissible.

\section{Squeezed vacuum}

As is discussed in the Introduction, when resonant atoms are inserted into
a medium with well developed polariton effects, then this medium can form
an effective squeezed vacuum [22--28]. Gardiner [20] noticed that spontaneous
emission of a single atom can be essentially distorted by the influence of
squeezed vacuum. Here we study how squeezed vacuum influences the attenuation
of an ensemble of many radiating atoms.

An effective squeezed vacuum appears due to the interaction of electromagnetic
field with matter. Local fluctuations of this effective squeezed vacuum are
superimposed on the local fluctuations caused by dipolar atomic interactions.
The latter are well represented by white noise. The effective squeezed vacuum
is usually modelled by parametric oscillators [20--23]. Therefore the
fluctuating field (42) can be written as a sum of two terms:
\be
\label{76}
\xi(t) = \xi_0(t) + \sum_q \gm(\om_q)\left ( b_q e^{-i\om_q t} +
b_q^\dgr e^{i\om_q t} \right ) \; .
\ee
Here the first term corresponds to white noise, for which
\be
\label{77}
\ll \xi_0(t)\gg\; = 0 \; , \qquad
\ll \xi_0^*(t)\xi_0(t')\gg\; = 2\gm_3\dlt(t-t') \; ,
\ee
analogously to Eq. (47). The second term in Eq. (76) describes a system of
oscillators modelling an effective squeezed vacuum. The creation, $b_q^\dgr$,
and annihilation, $b_q$, operators satisfy the Bose commutation relations. The
spectrum $\om_q=\om_{-q}$, in which $q$ is momentum, is positive. The spectral
function $\gm(\om)$ is assumed to be symmetric with respect to the central line
$\om_s$, so that
$$
\gm(\om_s +\om_q) = \gm(\om_s-\om_q) \; .
$$
The averages of the oscillator operators give
\be
\label{78}
\ll b_q\gg \; = 0 \; , \qquad \ll b_q^\dgr b_p\gg \; = n_q\dlt_{qp} \; ,
\qquad \ll b_q b_p^\dgr\gg \; = (1+n_q)\dlt_{qp} \; ,
\ee
where $n_q$ is a momentum distribution function. The averages $\ll b_qb_p\gg$
and $\ll b_q^\dgr b_p^\dgr\gg$ in the case of normal vacuum are zero, while for
a squeezed vacuum such anomalous averages do not vanish,
\be
\label{79}
\ll b_q^\dgr b_p^\dgr\gg \; = m_q\Dlt(\om_q+\om_p-2\om_s) \; ,
\ee
where $\Dlt(\om)$ is the discrete delta-function
\begin{eqnarray}
\Dlt(\om)\equiv \left\{ \begin{array}{cc}
1, & \om =0  \\
\nonumber
0, & \om \neq 0
\end{array}\right.
\end{eqnarray}
and $m_q$ is prescribed by the properties of squeezed vacuum. The momentum
distribution $n_q$ is semipositive, but the anomalous factor $m_q$ is, in
general, complex,
\be
\label{80}
n_q \equiv n(\om_q) \geq 0 \; , \qquad m_q \equiv |m(\om_q)| e^{i\vp_s} \; ,
\ee
with the phase $\vp_s$ defined by the squeezing field proportional to
$\cos(2\om_st+\vp_s)$. The white-noise and squeezed-vacuum fluctuations are
not correlated with each other, implying that
$$
\ll \xi_0(t) b_q\gg \; = 0 \; .
$$
Characteristic expressions for $n_q$ and $m_q$ are given in Appendix.

For the correlator of the fluctuating field (76), we find
$$
\ll \xi^+(t)\xi(t')\gg\; = 2\gm_3\dlt(t-t') +
\sum_q \gm^2(\om_q) \left [ n_q e^{i\om_q(t-t')} + \right.
$$
\be
\label{81}
\left.  + (1+n_q)e^{-i\om_q(t-t')}
+ m_q e^{i\om_qt+i(2\om_s-\om_q)t'} + m_q^* e^{-i\om_qt-i(2\om_s-\om_q)t'}
\right ] \; .
\ee
Substituting this in Eq. (44), we obtain the quantum attenuation as a sum
\be
\label{82}
\Gm_3 = \gm_3 + \Gm_{res} + \Gm_{non} + \Gm_s \; .
\ee
Here $\gm_3$ is due the white noise caused by dipolar interactions. The
resonant part of the quantum attenuation, due to the effective vacuum
fluctuations, is
\be
\label{83}
\Gm_{res} = n(\om) \; \frac{\gm^2(\om)\Gm}{\dlt^2+\Gm^2}\;
\left ( 1 - e^{-\Gm t}\right ) \; ,
\ee
where $\om$ is the frequency of the seed field, $\Gm$ and $\dlt$ are defined
in Eqs. (37) and (39), respectively, and the detuning from resonance is assumed
to be small, $|\dlt|\ll\om$. The nonresonant attenuation writes as
\be
\label{84}
\Gm_{non} = \Gm\sum_{\om_q\neq\om} \gm^2(\om_q) \; \left [
\frac{n_q}{(\Om-\om_q)^2+\Gm^2} + \frac{1+n_q}{(\Om+\om_q)^2+\Gm^2}
\right ] \; .
\ee
And the last term in Eq. (82) is the attenuation generated by squeezing,
\be
\label{85}
\Gm_s=-|m(\om)|\; \gm^2(\om)\;
\frac{\Gm\cos\vp_s+2\om_s\sin\vp_s}{4\om_s^2+\Gm^2} \; e^{-\Gm t} \; .
\ee

At the initial time $t=0$, the resonant attenuation (83), as is seen,
vanishes. Hence, the start of the radiation process is mainly due to the
noise attenuation $\gm_3$, nonresonant attenuation (84) and the squeezing
attenuation (85). For large times, the resonant attenuation (83) becomes more
important than nonresonant quantities (84) and (85). In this way, the
presence of squeezed vacuum results in the change of the attenuation
$\gm_3$ to an effective attenuation (82). Changing the attenuation leads
to the shift of the delay time (53), defining the maximum of the first
coherent burst, and also to the variation of the stationary solutions (61)
to (65). Consequently, there would arise some modifications in the temporal
behaviour of atomic squeezing characterized by the squeezing factor (66).
The overall evolution of the latter remains qualitatively the same. But
the details of this evolution can be regulated by the properties of
squeezed vacuum.

\section{Conclusion}

We have considered the atomic squeezing for an ensemble of many resonant
atoms. The used approach makes it possible to analyze all stages of
emission, which are: interaction stage (48), quantum stage (49), coherent
stage (55), relaxation stage (57), and quasistationary stage (60). The
temporal behaviour of the squeezing factor is studied in the course of its
evolution through all these stages. Both analytical estimates and numerical
calculations are presented. We show how it is possible to govern the
behaviour of the squeezing factor in the regime of punctuated superradiance.
This opens the way of an effective regulation of atomic squeezing by 
suppressing the fluctuations in the population difference with respect to
the fluctuations of the transition dipoles.

In the media with a well developed polariton effect, an effective squeezed
vacuum can arise. By incorporating resonant atoms into such media, one may
change the characteristics of the atomic system. The influence of the
squeezed vacuum on the collective radiation by atoms consists in modifying
the temporal characteristics of emission and varying the stationary solutions.
Hence, by preparing different kinds of squeezed vacuum, it is feasible to
regulate collective atomic emission and the evolution of atomic squeezing.

\vskip 5mm

{\bf Acknowledgement}

\vskip 2mm

We are very grateful to R. Tana\'s for many useful discussions and helpful 
advice. Financial support from the Bogolubov-Infeld Program and the German
Research Foundation (DFG grant Be 142/72-1) is appreciated. 

\newpage

{\large{\bf Appendix}}

\vskip 5mm

To present  explicit examples of the momentum distribution $n(w)$ and of
the anomalous term $|m(\om)|$ in the case of an effective squeezed vacuum,
we can resort to the standard case of a parametric oscillator [20--22], with
a squeezing field of intensity $\ep$ (in dimensionless units). One employs
the notation
$$
\mu\equiv \gm-\ep \; , \qquad \nu\equiv \gm+\ep \; ,
$$
with $\gm$ being a cavity damping rate. Then, for a nondegenerate parametric
oscillator, one has
$$
n(\om) = \frac{\nu^2-\mu^2}{8}\; \left [ \frac{1}{(\Dlt_s+\kappa)^2+\mu^2}
+\frac{1}{(\Dlt_s-\kappa)^2+\mu^2}\; - \; \frac{1}{(\Dlt_s+\kappa)^2+\nu^2}
\; - \; \frac{1}{(\Dlt_s-\kappa)^2+\nu^2} \right ] \; ,
$$
$$
|m(\om)| = \frac{\nu^2-\mu^2}{8}\; \left [ \frac{1}{(\Dlt_s+\kappa)^2+\mu^2}
+ \frac{1}{(\Dlt_s-\kappa^2)^2+\mu^2} + \frac{1}{(\Dlt_s+\kappa)^2+\nu^2}+
\frac{1}{(\Dlt_s-\kappa)^2+\nu^2} \right ] \; ,
$$
where $\Dlt_s\equiv\om-\om_s$ and the parameter $\kappa$ characterizes
a two-mode squeezed field, representing the displacement from the central
frequency of squeezing, where the two-mode squeezed vacuum is maximally
squeezed. If the parametric oscillator is weakly nondegenerate, such that
$|\kappa|\ll|\Dlt_s|$, or degenerate, with $\kappa=0$, then
$$
n(\om) = \frac{\nu^2-\mu^2}{4}\; \left ( \frac{1}{\Dlt_s^2+\mu^2}\; - \;
\frac{1}{\Dlt_s^2+\nu^2} \right ) \; ,
$$
$$
|m(\om)| = \frac{\nu^2-\mu^2}{4}\; \left ( \frac{1}{\Dlt_s^2+\mu^2} +
\frac{1}{\Dlt_s^2+\nu^2} \right ) \; .
$$
As is evident, generally, the normal, $n(\om)$, and the anomalous, $|m(\om)|$,
terms are of the same order of magnitude.

\newpage

\begin{center}
{\large{\bf Figure Captions}}
\end{center}

{\bf Fig. 1}. Temporal behaviour of the coherence intensity $w$ (solid line)
and population difference $s$ (dashed line) for $\gm_1=0.1$, $\gm_3=1$ (in
units of $\gm_2$), and $g=100$, without pumping, so that $\zeta=-1$. Both $w$
and $s$ are dimensionless. Time is measured in units of $\gm_2^{-1}$.

\vskip 5mm

{\bf Fig. 2}. Evolution of the squeezing factor $Q$ for the parameters of
Fig. 1. Time is measured in units of $\gm_2^{-1}$.

\vskip 5mm

{\bf Fig. 3}. Squeezing factor $Q$ for the parameters $\gm_1=\gm_3=1$,
$g=100$, and $\zeta=-1$. Time is measured in units of $\gm_2^{-1}$.

\vskip 5mm

{\bf Fig. 4} Oscillations of the coherence intensity $w$ and population
difference $s$ as functions of time (in units of $\gm_2^{-1}$) in the 
presence of pumping, when $\zeta=1$, for the parameters  $\gm_1=\gm_3=1$, 
$g=10$.

\vskip 5mm

{\bf Fig. 5} Evolution of the squeezing factor $Q$ for the parameters of
Fig. 4. Time is measured in units of $\gm_2^{-1}$.

\vskip 5mm

{\bf Fig. 6}. Regulating the number of oscillations of $w$ and $s$, under
$\zeta=1$, by varying the system parameters, which are $\gm_1=1,\;
\gm_3=0.1$, and $g=100$. Time is measured in units of $\gm_2^{-1}$.

\vskip 5mm

{\bf Fig. 7} Evolution of the squeezing factor $Q$ for the parameters of
Fig. 6. Time is measured in units of $\gm_2^{-1}$.

\vskip 5mm

{\bf Fig. 8} Governing the behaviour of the squeezing factor $Q$ in the
regime of punctuated atomic superradiance for the parameters $\gm_1=0.1,\;
\gm_3=1$, $g=100$, and $\zeta=-1$, with equidistant inverting pulses at the
moments of time $t_1=0.1$, $t_2=0.2$, $t_3=0.3$, $t_4=0.4$, and $t_5=0.5$.
Time is measured in units of $\gm_2^{-1}$.

\vskip 5mm

{\bf Fig. 9} Governing the evolution of the squeezing factor $Q$ in the
regime of punctuated atomic superradiance for the same parameters as in
Fig. 8, but with nonequidistant inverting pulses at the moments of time
$t_1=0.1$, $t_2=0.2$, $t_3=0.3$, $t_4=0.4$, $t_5=0.5$, $t_6=1$, $t_7=1.1$,
$t_8=1.2$, $t_9=2$, $t_{10}=2.4$, and $t_{11}=2.5$. Time is measured in 
units of $\gm_2^{-1}$.

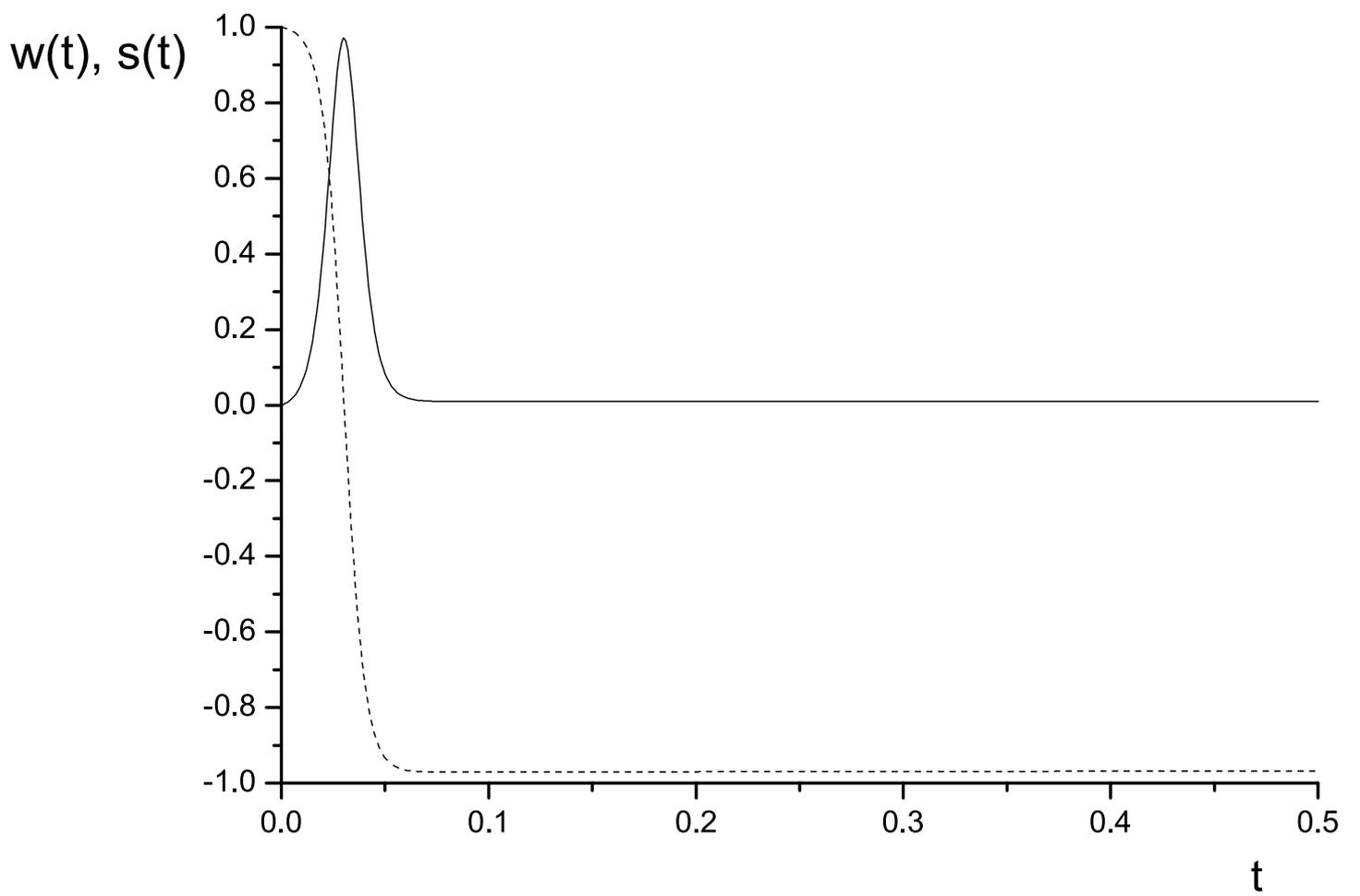
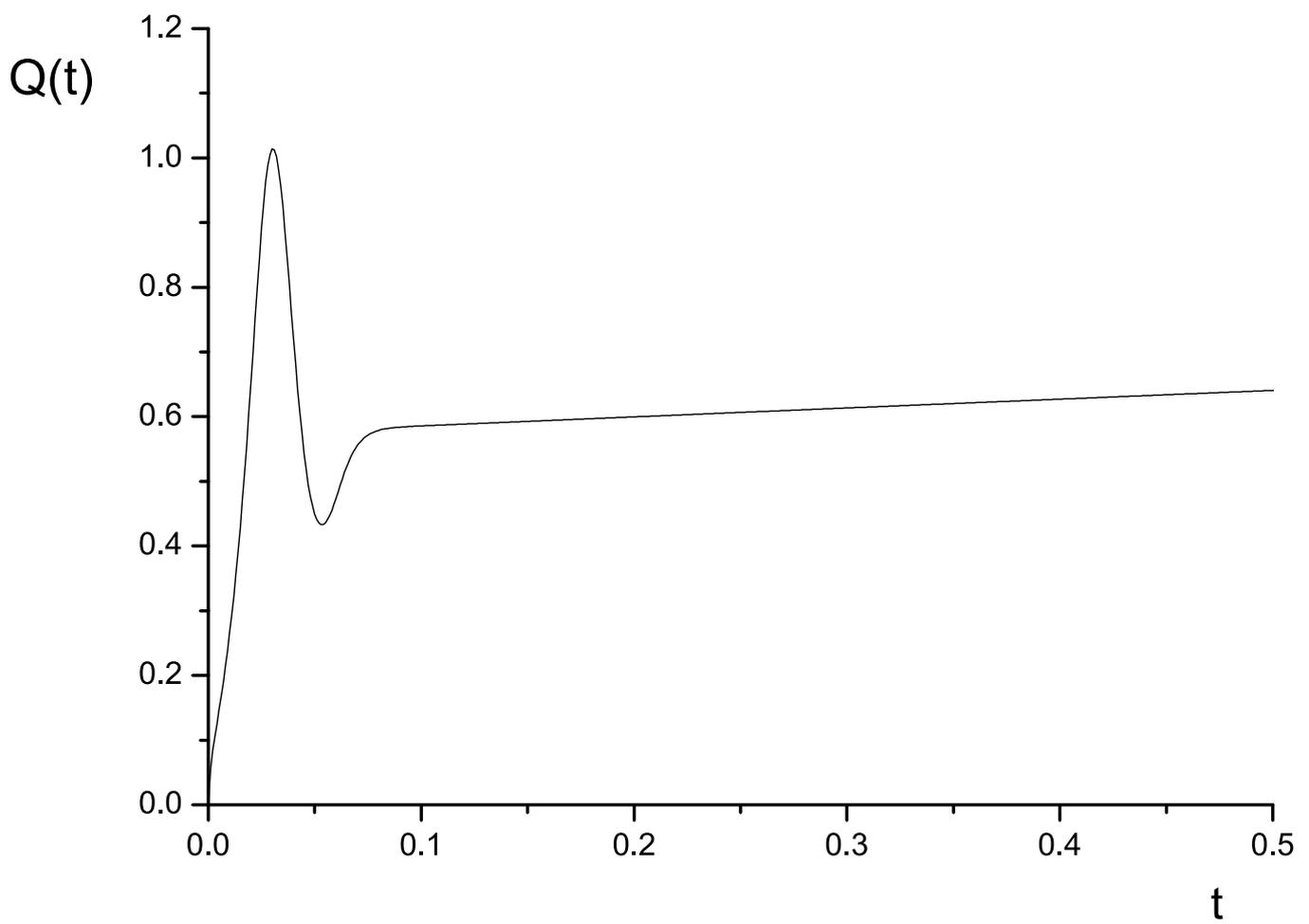
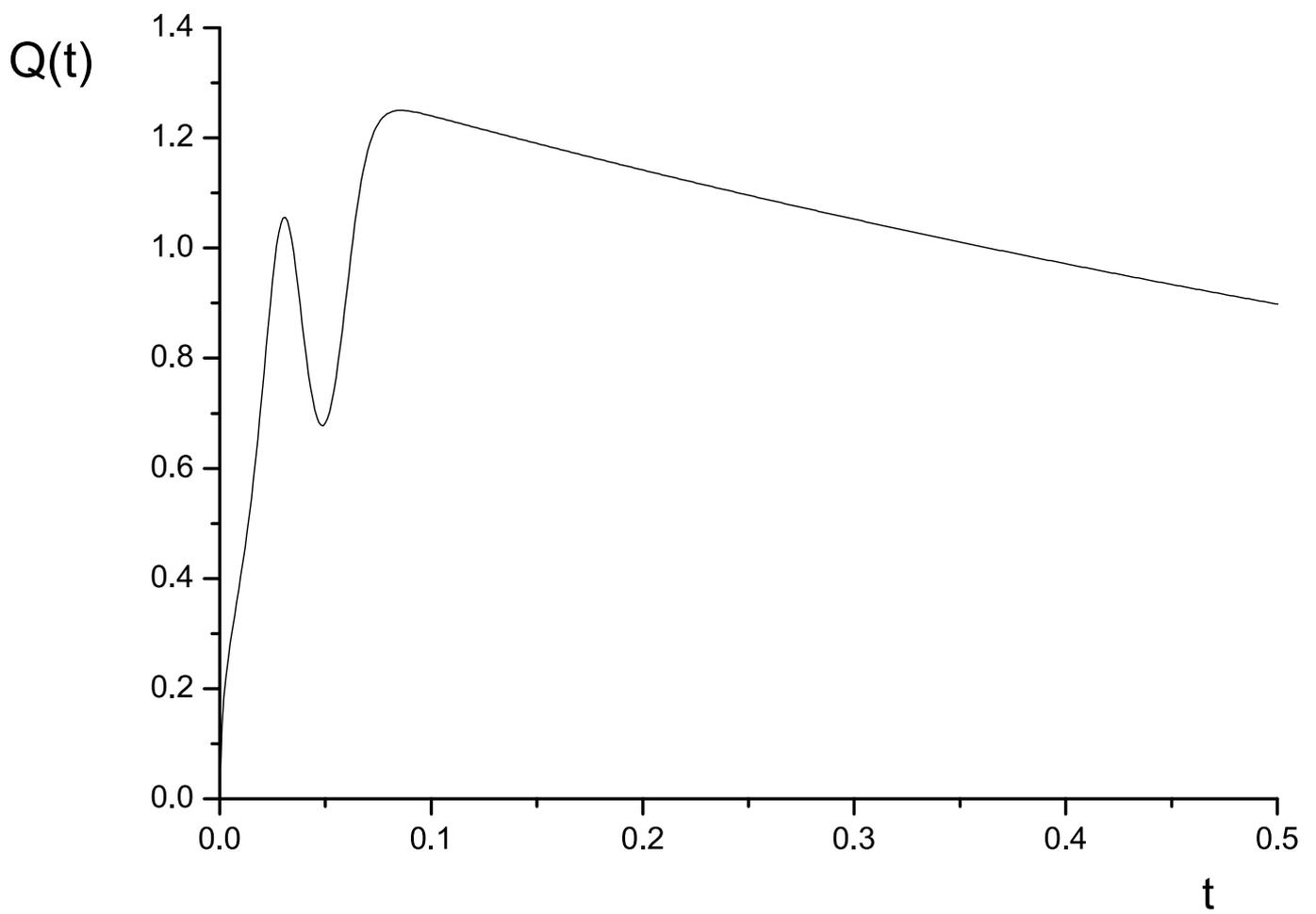
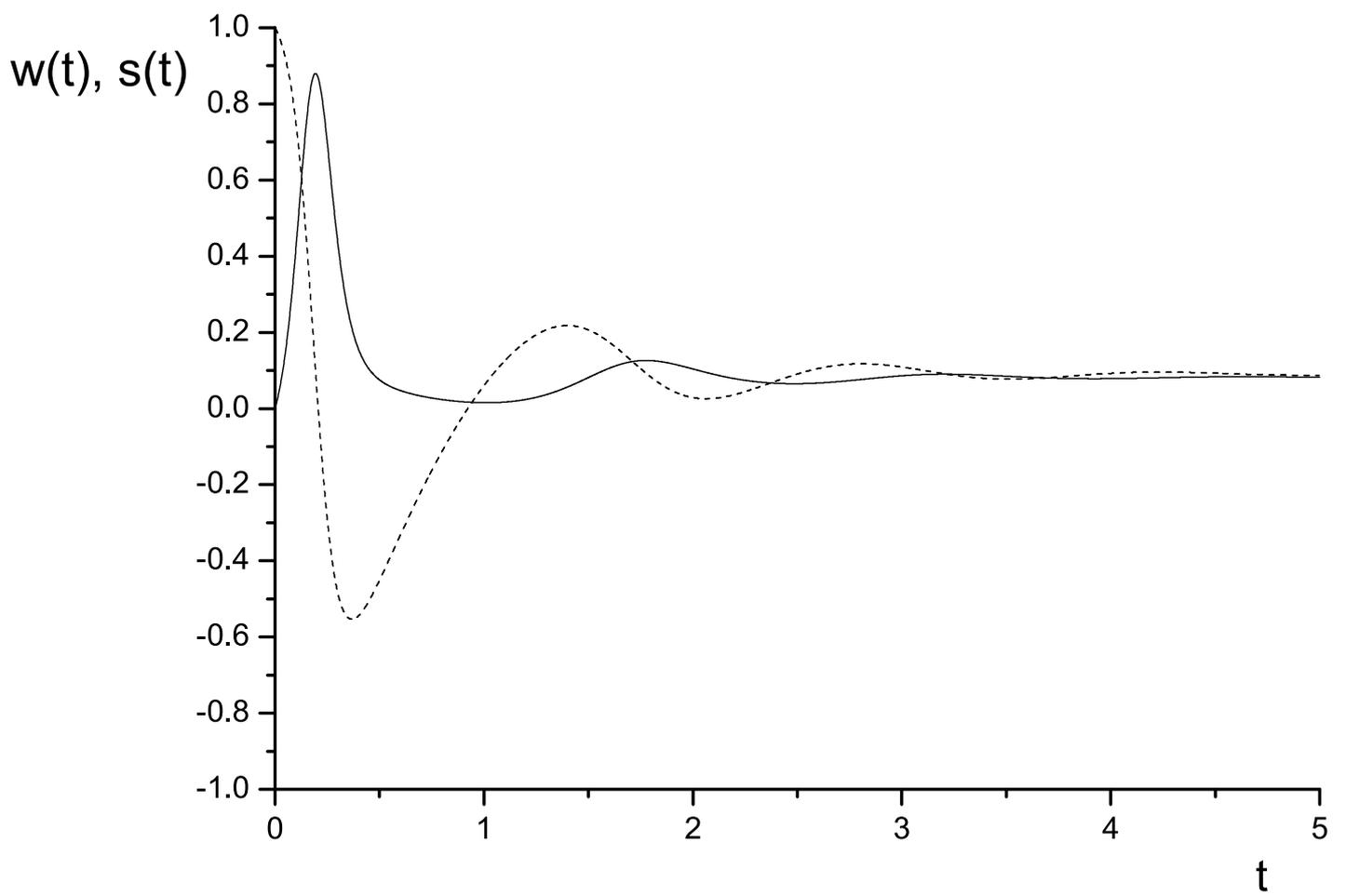
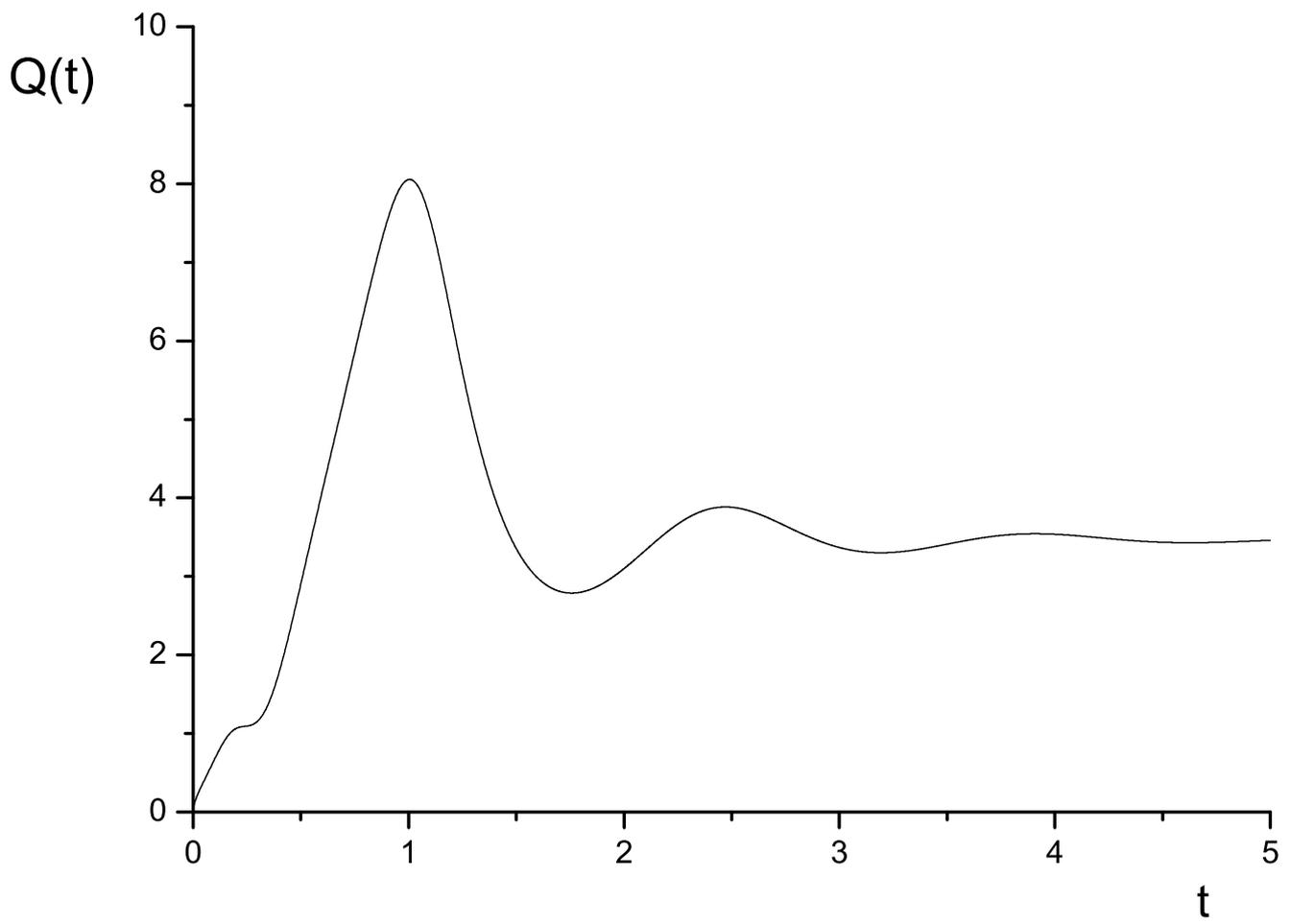
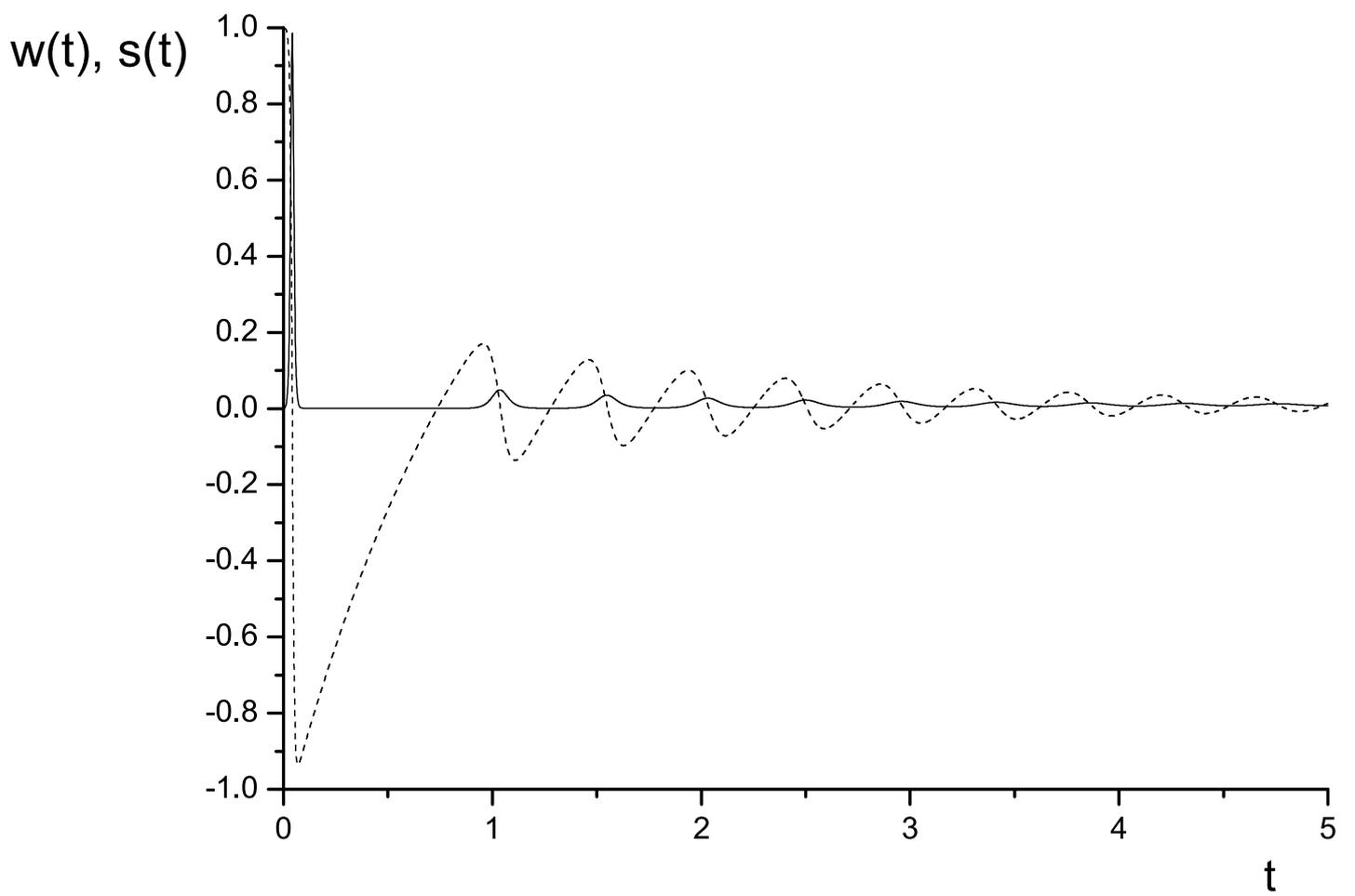
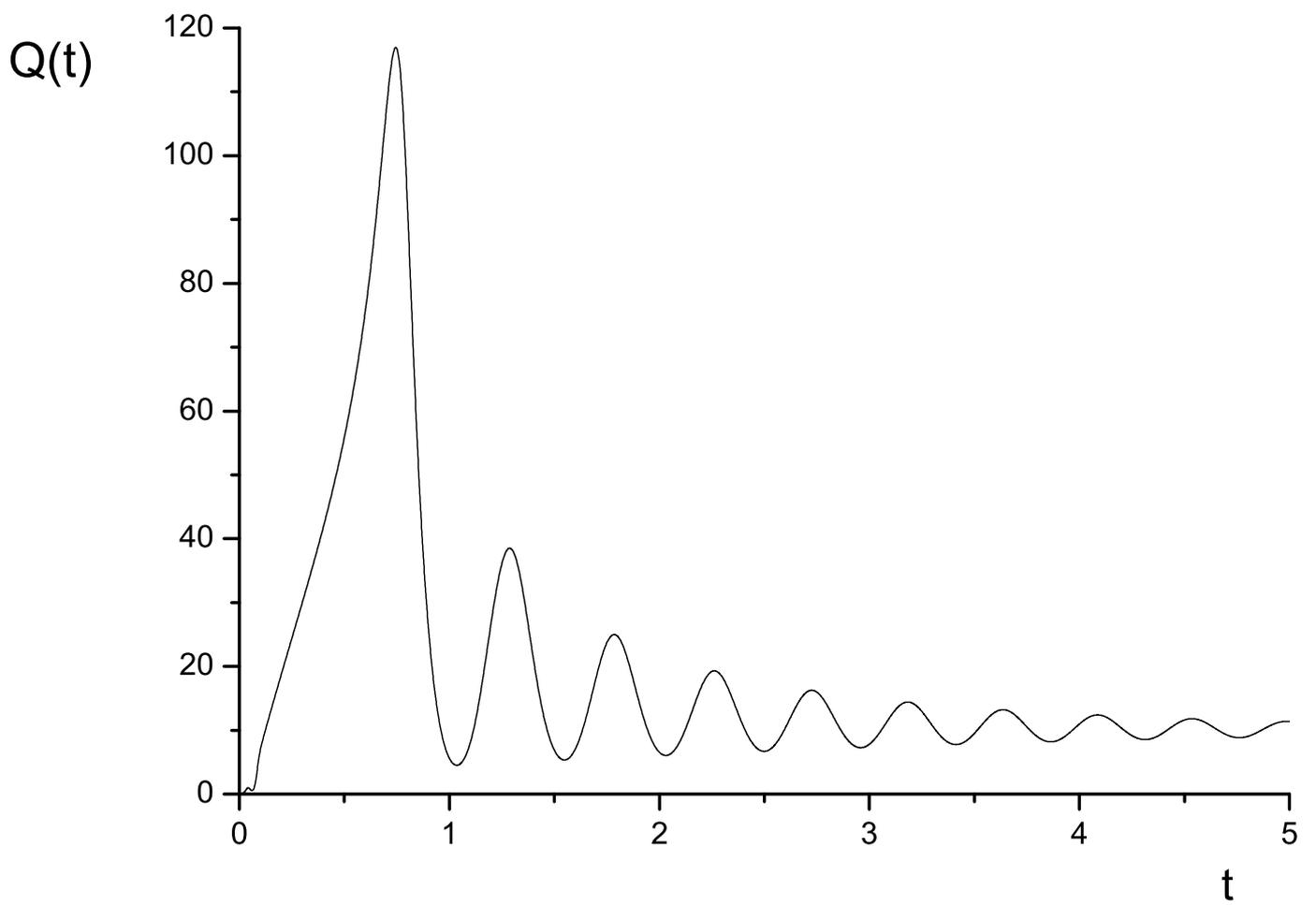
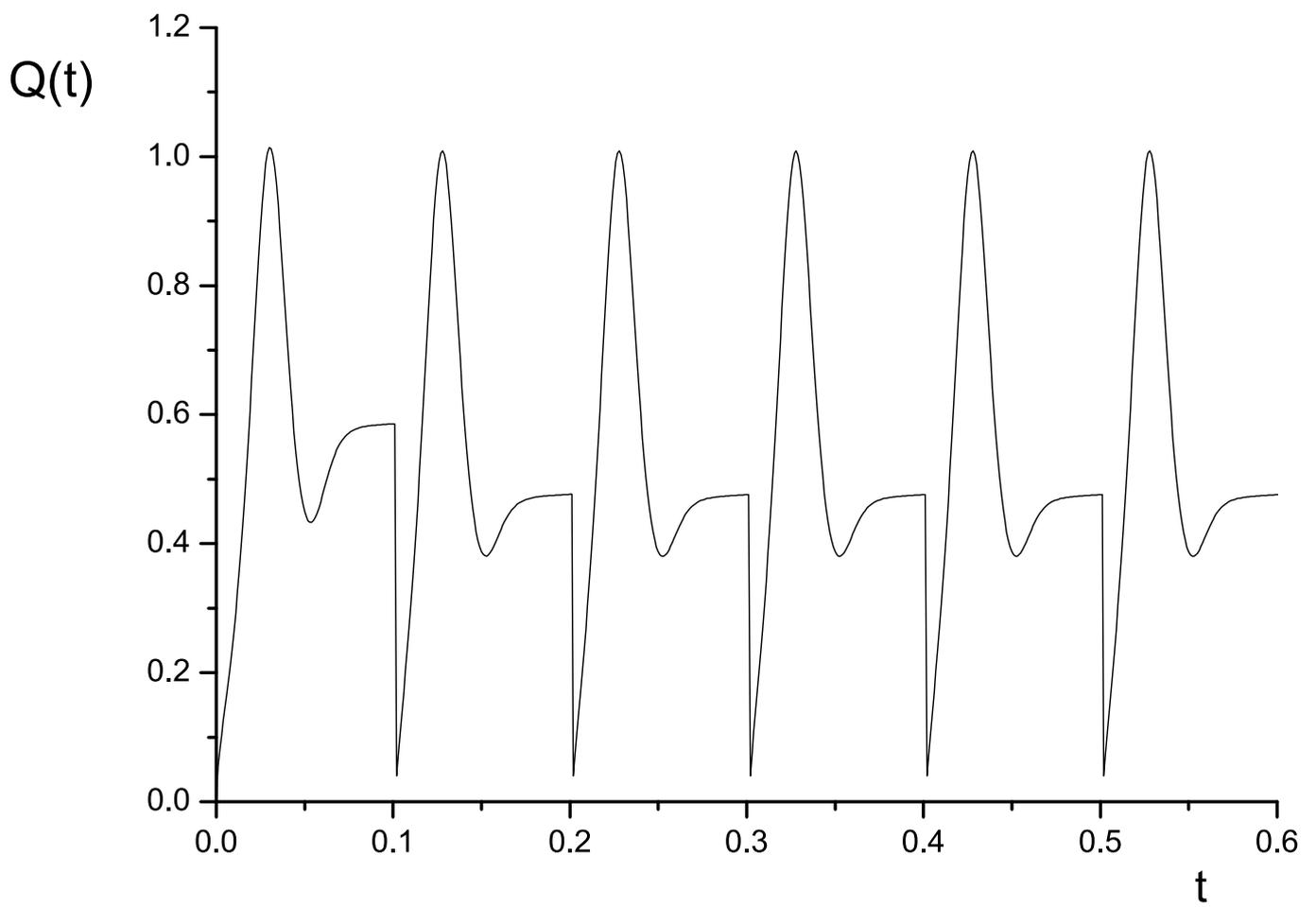
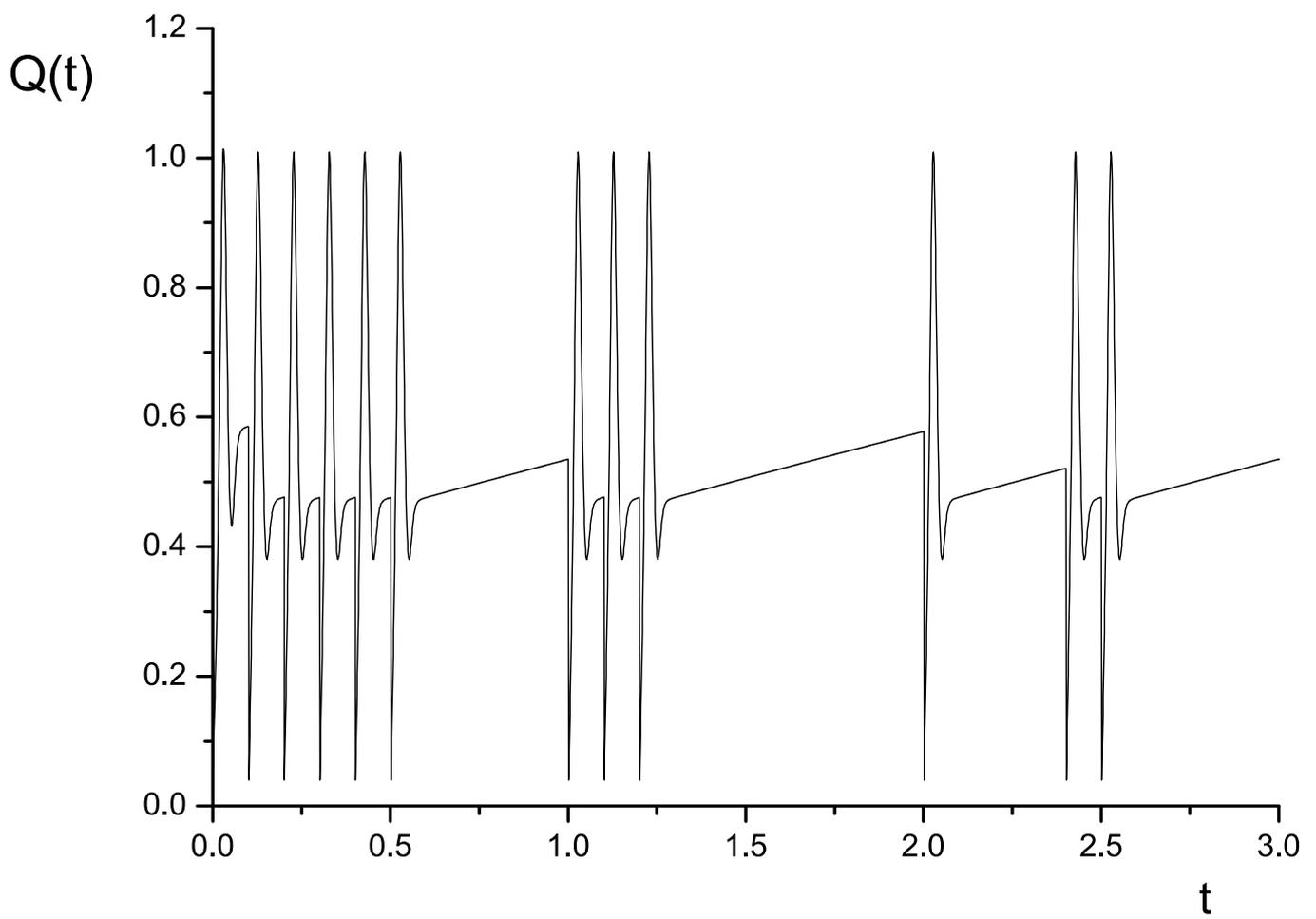
\end{document}